----------------------------------------------------------------
\documentstyle[aps,12pt]{revtex}
\begin{document}

\title{Quantum three-body system in $D$ dimensions}

\author{Xiao-Yan Gu $^{a,b,c}$ \thanks{Electronic address:
guxy@mail.ihep.ac.cn}, Bin Duan $^{a,b}$, and Zhong-Qi Ma $^{a,b,c}$}

\address{a) CCAST (World Laboratory), P.O.Box 8730, Beijing 100080, China \\
b) Institute of High Energy Physics, P.O.Box 918(4), Beijing 100039, China
\thanks{Mailing address}\\
c) Graduate School of Chinese Academy of Sciences, Beijing 100039, China}


\maketitle

\vspace{5mm}

\begin{abstract}
The independent eigenstates of the total orbital angular 
momentum operators for a three-body system in an arbitrary 
$D$-dimensional space are presented by the method of group theory. 
The Schr\"{o}dinger equation is reduced to the generalized 
radial equations satisfied by the generalized radial functions 
with a given total orbital angular momentum denoted by a Young diagram
$[\mu,\nu,0,\ldots,0]$ for the SO$(D)$ group. Only three internal 
variables are involved in the functions and equations. The number 
of both the functions and the equations for the given angular 
momentum is finite and equal to $\left(\mu-\nu+1\right)$.  

\end{abstract}

\section{Introduction}

From the very early stage of the progress in quantum mechanics 
in the real three-dimensional world, it has been pointed out that 
the essence of these theories would be easily understandable if 
their mathematics is constructed in the non-relativistic
hyperspace worlds \cite{foc,bar}. The mathematical tools for
generalization of the orbital angular momentum in an arbitrary 
$D$-dimensional space have been presented \cite{lou1,lou2,lou3,cha,ban}. 
Recently, the $D$-dimensional Coulombic and the harmonic 
oscillator problems in a two-body system have been studied 
in some detail by many authors 
\cite{wod,ben1,ben2,rom,alj1,lin,hos,aqu1,aqu2,alj2,dab,nou,bur,gon,chan}. 

Exact solutions played very important roles in the development of
physics. The exact solutions of the Schr\"{o}dinger equation in
the real three dimensional space for a hydrogen atom and for a 
harmonic oscillator were important technical achievements in 
quantum mechanics \cite{sch}, which provided strong evidence in 
favor of the theory being correct, at least as far as atomic 
physics is concerned. The next simplest atom is the helium atom, 
for which the Schr\"{o}dinger equation cannot be solved analytically, 
but only numerically \cite{fan,kri,lin1,haf,tan}. In the numerical 
calculation, one of the main difficulties is how to separate 
the global rotational degrees of freedom. 

In our previous paper \cite{gu} we separated completely the 
global rotational degrees of freedom in the Schr\"{o}dinger 
equation for an $N$-body system in the real three-dimensional
space from the internal ones. We have determined a complete 
set of $(2l+1)$ independent base functions for a given total 
orbital angular momentum $l$, which are the homogeneous 
polynomials in the components of coordinate vectors and do
not contain the Euler angles explicitly. Any function with 
the given angular momentum $l$ in the system can be expanded 
with respect to the base functions, where the coefficients
are the functions of the internal variables, called the 
generalized radial functions. The generalized radial equations 
satisfied by the functions are established explicitly \cite{gu}. 
For the typical three-body system in the real three dimensional 
space \cite{man1,man2}, such as a helium atom \cite{duan1,duan2} 
and a positronium negative ion \cite{duan3}, the generalized 
radial equations \cite{ma1} have been solved numerically with 
high precision. 

With the interest of higher dimensional field theory recently, 
we have a try to generalize the study of the $D$-dimensional
two-body system to the $D$-dimensional three-body system. The 
purpose of this paper is, for a three-body system in an arbitrary 
$D$-dimensional space, to find a complete set of independent 
base functions with any given total orbital angular momentum 
and to reduce the Schr\"{o}dinger equation with a spherically 
symmetric potential $V$ to the generalized radial equations, 
where only three internal variables are involved. Any function 
with the given angular momentum in the system can be expanded 
with respect to the base functions. It provides a possibility
to calculate numerically the energy levels of the three-body 
system in $D$-dimensions with high precision. 

From the viewpoint of mathematics, the separation of the global 
rotational degrees of freedom from the internal ones is a 
typical application of group theory to physics. The properties 
of the independent base functions for a given total orbital 
angular momentum would be more clear if they are constructed 
in arbitrary $D$-dimensional space than that in the real three 
dimensional space. The total orbital angular momentum for a 
three-body system in a $D$-dimensional space is described by 
an irreducible representation denoted by a Young diagram with one or
two rows. For the real three-dimensional space, the rotational 
symmetry group is SO(3) group, and its only irreducible 
representations denoted by the Young diagrams with two rows are $[l,1]$, 
which are equivalent to the representations denoted by the one-row 
Young diagrams $[l,0]$, respectively. This is the reason why the 
angular momentum can be described by only one quantum number $l$ 
for the real three-dimensional space. 

This paper is organized as follows. After separating the motion
of the center of mass by the Jacobi coordinate vectors in Sec. 2, 
we review in Sec. 3 the generalization of the orbital angular 
momentum operators and the properties of the spherical harmonics 
\cite{lou2,cha} and the harmonic polynomials \cite{ban} for a 
two-body system in $D$ dimensions. In Sec. 4 we will define the 
generalized harmonic polynomials for a three-body system in $D$ 
dimensions and prove that they constitute a complete set of independent 
base functions for a given total orbital angular momentum in the system. 
The generalized radial functions are defined and the generalized 
radial equations are derived in Sec. 5. Some conclusions will be 
given in Sec. 6.

\section{Schr\"{o}dinger equation in $D$ dimensions}

For a quantum $N$-body system in an arbitrary $D$-dimensional space, 
we denote the position vectors and the masses of $N$ particles by 
${\bf r}_{k}$ and by $m_{k}$, $k=1,2,\ldots,N$, respectively. 
$M=\sum_{k} m_{k}$ is the total mass. The Schr\"{o}dinger equation 
for the $N$-body system with a pair potential $V$, depending upon 
the distance of each pair of particles, $|{\bf r}_{j}-{\bf r}_{k}|$,
is 
$$- \displaystyle {1 \over 2} \displaystyle 
\sum_{k=1}^{N}~\displaystyle m_{k}^{-1} 
\bigtriangledown^{2}_{{\bf r}_{k}} \Psi 
+V \Psi =E \Psi , \eqno (1) $$

\noindent
where $\bigtriangledown^{2}_{{\bf r}_{k}}$ is the Laplace operator with 
respect to the position vector ${\bf r}_{k}$. For simplicity, the 
natural units $\hbar=c=1$ are employed throughout this paper.
The total orbital angular momentum operators $L_{ab}$ in $D$ dimensions 
are defined as \cite{lou2,cha} 
$$L_{ab}=-L_{ba}=-i\displaystyle \sum_{k=1}^{N}~\left\{
r_{ka}\displaystyle {\partial \over \partial r_{kb}}
-r_{kb}\displaystyle {\partial \over \partial r_{ka}} \right\},~~~~~~
a,b=1,2,\ldots D, \eqno (2) $$

\noindent
where $r_{ka}$ denotes the $a$th component of the position vector
${\bf r}_{k}$. 

Now, we replace the position vectors ${\bf r}_{k}$ by the Jacobi
coordinate vectors ${\bf R}_{j}$:
$${\bf R}_{0}=M^{-1/2}\displaystyle \sum_{k=1}^{N}~m_{k}{\bf r}_{k},~~~~~
{\bf R}_{j}=\left(\displaystyle {m_{j+1}M_{j}\over M_{j+1}}
 \right)^{1/2}\left({\bf r}_{j+1}-\displaystyle \sum_{k=1}^{j}~
\displaystyle {m_{k}{\bf r}_{k}\over M_{j}}\right), $$
$$1\leq j \leq (N-1),~~~~~~M_{j}=\displaystyle 
\sum_{k=1}^{j}~m_{k},~~~~~~M_{N}=M,  \eqno (3) $$

\noindent
where ${\bf R}_{0}$ describes the position of the center of mass,
${\bf R}_{1}$ describes the mass-weighted separation from the second 
particle to the first particle. ${\bf R}_{2}$ describes the mass-weighted 
separation from the third particle to the center of mass of the first 
two particles, and so on. An additional factor $\sqrt{M}$ 
is included in ${\bf R}_{j}$ for convenience. The mass-weighted 
factors in front of the formulas for ${\bf R}_{j}$ are determined 
by the condition
$$\sum_{k=1}^{N}~m_{k}{\bf r}_{k}^{2}
=\sum_{j=0}^{N-1}~{\bf R}_{j}^{2}, $$ 

\noindent
One may determine the factors one by one from the following 
schemes. In the center-of-mass frame, if the first 
$j$ particles coincide with each other and the last $(N-j-1)$
particles are located at the origin, the factor in front of 
${\bf R}_{j}$ is determined by
$${\bf r}_{1}={\bf r}_{2}=\cdots ={\bf r}_{j}
=-m_{j+1}{\bf r}_{j+1}/M_{j},~~~~~
\displaystyle \sum_{k=1}^{j+1}~m_{k}{\bf r}_{k}^{2}
={\bf R}_{j}^{2}. \eqno (4) $$

A straightforward calculation by replacement of variables shows 
that the Laplace operator in Eq. (1) and the total orbital angular 
momentum operator $L_{ab}$ in Eq. (2) are directly expressed in 
${\bf R}_{j}$:
$$\begin{array}{l}
\bigtriangledown^{2} =\displaystyle \sum_{k=1}^{N}~\displaystyle
m_{k}^{-1} \bigtriangledown^{2}_{{\bf r}_{k}}
= \displaystyle \sum_{j=0}^{N-1}~
\bigtriangledown^{2}_{{\bf R}_{j}}, \\[2mm]
L_{ab}=-i\displaystyle \sum_{j=0}^{N-1}~\left\{
R_{ja}\displaystyle {\partial \over \partial R_{jb}}
-R_{jb}\displaystyle {\partial \over \partial R_{ja}} \right\}. 
\end{array} \eqno (5) $$

In the center-of-mass frame, ${\bf R}_{0}=0$. The Laplace 
operator (5) obviously has the symmetry of the O$(ND-D)$ group 
with respect to $(N-1)D$ components of $(N-1)$ Jacobi coordinate 
vectors. The O$(ND-D)$ group contains a subgroup SO($D)\times$O$(N-1)$, 
where SO$(D)$ is the rotation group in the $D$-dimensional space. 
The space inversion and the different definitions for the Jacobi 
coordinate vectors in the so-called Jacobi tree \cite{fan} can 
be obtained by O$(N-1)$ transformations. For the system of 
identical particles, the permutation group among particles is 
also a subgroup of the O$(N-1)$ group \cite{gu}. 

It is easy to obtain the inverse transformation of Eq. (3):
$${\bf r}_{j}=\left[M_{j-1} \over m_{j}M_{j}\right]^{1/2}{\bf R}_{j-1}
-\displaystyle \sum_{k=j}^{N-1}~
\left[m_{k+1} \over M_{k}M_{k+1}\right]^{1/2}{\bf R}_{k}
+M^{-1/2}{\bf R}_{0},  $$
$${\bf r}_{j}-{\bf r}_{k}=
\left[M_{j} \over m_{j}M_{j-1}\right]^{1/2}{\bf R}_{j-1}
+\displaystyle \sum_{i=k}^{j-2}~
\left[m_{i+1} \over M_{i}M_{i+1}\right]^{1/2}{\bf R}_{i}
-\left[M_{k-1} \over m_{k}M_{k}\right]^{1/2}{\bf R}_{k-1}. \eqno (6) $$

\noindent
Thus, the potential $V$ is a function of 
${\bf R}_{j}\cdot {\bf R}_{k}$ and is rotationally invariant.

\section{Harmonic polynomials in $D$ dimensions}

In the center-of-mass frame, ${\bf R}_{0}=0$. Hence, for a 
two-body system there is only one Jacobi coordinate vector 
${\bf R}_{1}$, which will be denoted by ${\bf x}$ for simplicity:
$$\begin{array}{c}
{\bf x}=\left(\displaystyle {m_{1}m_{2} \over m_{1}+m_{2}}\right)^{1/2}
\left\{{\bf r}_{2}-{\bf r}_{1}\right\}, \\[2mm]
\bigtriangledown^{2}=\bigtriangledown^{2}_{\bf x},~~~~~~
L_{ab}=-i\left\{x_{a}\displaystyle {\partial \over \partial x_{b}}
-x_{b}\displaystyle {\partial \over \partial x_{a}} \right\}, 
\end{array} \eqno (7) $$

Louck \cite{lou2,cha} introduced the hyperspherical coordinates
$$\begin{array}{l}
x_{1}=r\cos \theta_{1} \sin \theta_{2} \ldots \sin \theta_{D-1},\\
x_{2}=r\sin \theta_{1} \sin \theta_{2} \ldots \sin \theta_{D-1}, \\
x_{k}=r\cos \theta_{k-1} \sin \theta_{k} \ldots \sin \theta_{D-1},~~~~~~
3\leq k \leq D-1,\\
x_{D}=r\cos \theta_{D-1}. \end{array} \eqno (8) $$

\noindent
The spherical harmonics $Y^{l}_{l_{D-2},\cdots, l_{1}}$ in 
$D$ dimensions \cite{lou2,cha} are the simultaneous eigenfunctions  
of the commutant operators ${\bf L}^{2}_{k}$:
$$\begin{array}{l}
{\bf L}_{1}^{2}=- \displaystyle 
   {\partial^{2} \over \partial \theta_{1}^{2}},~~~~~~
{\bf L}^{2}_{k}=- \left\{\displaystyle {1 \over 
\sin^{k-1} \theta_{k}}\displaystyle {\partial \over \partial \theta_{k}} 
\sin^{k-1} \theta_{k}{\partial \over \partial \theta_{k}}
-\displaystyle  { {\bf L}_{k-1}^{2} \over \sin^{2} \theta_{k} }
\right\},\end{array} \eqno (9) $$
$$\begin{array}{l}
{\bf L}_{1}^{2}Y^{l}_{l_{D-2},\cdots, l_{1}}(\theta_{1}\ldots \theta_{D-1})
=l_{1}^{2}Y^{l}_{l_{D-2},\cdots, l_{1}}(\theta_{1}\ldots \theta_{D-1}), \\
{\bf L}_{k}^{2}Y^{l}_{l_{D-2},\cdots, l_{1}}(\theta_{1}\ldots \theta_{D-1})
=l_{k}(l_{k}+k-1)Y^{l}_{l_{D-2},\cdots, l_{1}}(\theta_{1}\ldots \theta_{D-1}),
 \\
l\equiv l_{D-1}=0,1,\ldots,~~~~~l_{k}=0,1,\ldots,l_{k+1},~~~~~
l_{1}=-l_{2},-l_{2}+1,\ldots,l_{2}-1,l_{2}, \end{array} \eqno (10) $$

\noindent
where ${\bf L}^{2}\equiv {\bf L}_{D-1}^{2}$, $0\leq r < \infty$, 
$-\pi \leq \theta_{1} \leq \pi$, $0\leq \theta_{k} \leq \pi$, 
and $2\leq k \leq D-1$. The volume element of the configuration 
space is \cite{lou2,hos}
$$\displaystyle \prod_{j=1}^{D}~dx_{j}
=r^{D-1}dr \displaystyle \prod_{j=1}^{D-1}~
\left(\sin\theta_{j}\right)^{j-1}d\theta_{j}. \eqno (11) $$

\noindent
Through a direct calculation by replacement of variables, 
one obtains \cite{lou2,cha}
$$\bigtriangledown^{2}_{\bf x}=\displaystyle {1\over r^{D-1}}
\displaystyle {\partial \over \partial r} r^{D-1}
\displaystyle {\partial \over \partial r} 
-\displaystyle {{\bf L}^{2} \over r^{2}}, \eqno (12) $$

\noindent
Due to the spherical symmetry, the wave function can be expressed as
$$\psi^{l}_{l_{D-2},\cdots, l_{1}}({\bf x})=\phi_{l}(r)
Y^{l}_{l_{D-2},\cdots, l_{1}}(\theta_{1}\ldots \theta_{D-1}), \eqno (13) $$

\noindent
and the $D$-dimensional Schr\"{o}dinger equation (1) for a 
two-body system in the center-of-mass frame reduces to the 
radial equation
$$\displaystyle {1\over r^{D-1}}
\displaystyle {\partial \over \partial r} r^{D-1}
\displaystyle {\partial \over \partial r}\phi_{l}(r) 
-\displaystyle {l(l+D-2) \over r^{2}}\phi_{l}(r)
=-2\left[E-V(r)\right]\phi_{l}(r). \eqno (14) $$

Bander and Itzykson \cite{ban} introduced the harmonic polynomials
in $D$ dimensions 
$${\cal Y}^{l}_{l_{D-2},\cdots, l_{1}}({\bf x})
=r^{l}Y^{l}_{l_{D-2},\cdots, l_{1}}(\theta_{1}\ldots \theta_{D-1})
\equiv r^{l}Y^{l}_{l_{D-2},\cdots, l_{1}}(\hat{\bf x}), 
\eqno (15) $$

\noindent
to avoid the angular functions 
$Y^{l}_{l_{D-2},\cdots, l_{1}}(\theta_{1}\ldots \theta_{D-1})$. 
${\cal Y}^{l}_{l_{D-2},\cdots, l_{1}}({\bf x})$ is a homogeneous 
polynomial of degree $l$ in the components of ${\bf x}$ and 
satisfies the Laplace equation
$$\bigtriangledown^{2}_{\bf x} 
{\cal Y}^{l}_{l_{D-2},\cdots, l_{1}}({\bf x})=0. 
\eqno (16) $$

\noindent
The number of linearly independent homogeneous polynomials of 
degree $l$ in $D$ components of ${\bf x}$ is $N(l)=(l+D-1)!/l!(D-1)!$. 
The Laplace equation (16) gives $N(l-2)=(l+D-3)!/(l-2)!(D-1)!$ constraints. 
Hence, the number of the harmonic polynomials 
${\cal Y}^{l}_{l_{D-2},\cdots, l_{1}}({\bf x})$ of degree $l$ as 
well as the number of the spherical harmonics 
$Y^{l}_{l_{D-2},\cdots, l_{1}}(\hat{\bf x})$ in $D$ dimensions is 
$$N(l)-N(l-2)=\displaystyle {(2l+D-2)(l+D-3)!\over l!(D-2)!}
=d_{D}([l,0,\ldots,0]). \eqno (17) $$

\noindent
$d_{D}([l,0,\ldots,0])$ is the dimension of the irreducible 
representation of SO$(D)$ denoted by the one-row Young diagram 
$[l,0,\ldots,0]$. $[l,0,\ldots,0]$ describes the symmetric 
traceless tensor representation. In fact, any polynomial in 
the components of one vector ${\bf x}$ has to belong to a 
symmetric representation.            

Due to the spherical symmetry, one only needs to write the 
explicit form of the highest weight state \cite{ban}
$${\cal Y}^{l}_{l,\cdots, l}({\bf x})=N_{l}(x_{1}+ix_{2})^{l}. \eqno (18) $$

\noindent
where $N_{l}$ denotes the normalization factor. The partners of 
${\cal Y}^{l}_{l,\cdots, l}({\bf x})$ can be simply generated by 
rotation. Now, the solution to the Schr\"{o}dinger equation in 
the center-of-mass frame can be re-expressed as 
$$\psi^{l}_{l,\cdots,l}({\bf x})
=R_{l}(r){\cal Y}^{l}_{l,\cdots, l}({\bf x}), \eqno (19) $$

\noindent
and the radial equation is easy to be derived: 
$$\displaystyle {1\over r^{D-1}}
\displaystyle {\partial \over \partial r} r^{D-1}
\displaystyle {\partial \over \partial r}R_{l}(r)
+\displaystyle {2l \over r}\displaystyle {\partial \over \partial r} R_{l}(r)
=-2\left[E-V(r)\right]R_{l}(r). \eqno (20) $$

\noindent
Recall $R_{l}(r)=r^{-l}\phi_{l}(r)$. Eq. (20) coincides with Eq. (14) but
the angle variables do not appear explicitly in calculation. 

The number (17) of the harmonic polynomials
${\cal Y}^{l}_{l_{D-2},\cdots, l_{1}}({\bf x})$ of degree $l$ can 
be understood from another viewpoint. After removing those 
homogeneous polynomials in the form $r^{2}f({\bf x})$, where 
$f({\bf x})$ is a homogeneous polynomial of degree $(l-2)$, 
Eq. (17) shows the number of the remaining linearly 
independent homogeneous polynomials of degree 
$l$ in the components of ${\bf x}$. Therefore, the harmonic 
polynomials ${\cal Y}^{l}_{l_{D-2},\cdots, l_{1}}({\bf x})$ construct 
a complete set of linearly independent base functions for 
the homogeneous polynomials of degree $l$ in the components of 
${\bf x}$, excluded those in the form of $r^{2}f({\bf x})$. 

\section{Three-body system in $D$-dimensions}

For a three-body system, in the center-of-mass frame there are 
two Jacobi coordinate vectors ${\bf R}_{1}$ and ${\bf R}_{2}$, 
which will be denoted by ${\bf x}$ and ${\bf y}$, respectively:
$$\begin{array}{l}
{\bf x}=\left[\displaystyle {m_{1}m_{2} \over m_{1}+m_{2}}\right]^{1/2}
\left\{{\bf r}_{2}-{\bf r}_{1}\right\},~~~~~~
{\bf y}=\left[\displaystyle {(m_{1}+m_{2})m_{3} \over m_{1}+m_{2}+m_{3}}
\right]^{1/2}\left\{{\bf r}_{3}-\displaystyle 
{m_{1}{\bf r}_{1}+m_{2}{\bf r}_{2}\over m_{1}+m_{2}}\right\}, \\[2mm]
\end{array} $$
$$\begin{array}{l}
\bigtriangledown^{2}=\bigtriangledown^{2}_{\bf x}
+\bigtriangledown^{2}_{\bf y},\\
L_{ab}=L^{(x)}_{ab}+L^{(y)}_{ab}
=-i\left\{x_{a}\displaystyle {\partial \over \partial x_{b}}
-x_{b}\displaystyle {\partial \over \partial x_{a}} \right\}
-i\left\{y_{a}\displaystyle {\partial \over \partial y_{b}}
-y_{b}\displaystyle {\partial \over \partial y_{a}} \right\}, 
\end{array} \eqno (21) $$

\noindent
The Schr\"{o}dinger equation (1) reduces to 
$$\begin{array}{c}
\left\{\bigtriangledown^{2}_{\bf x}
+\bigtriangledown^{2}_{\bf y}\right\}\Psi({\bf x,y})
=-2\left\{E-V(\xi_{1},\xi_{2},\xi_{3})\right\}\Psi({\bf x,y}),\\
\xi_{1}={\bf x\cdot x},~~~~~~
\xi_{2}={\bf y\cdot y},~~~~~~\xi_{3}={\bf x\cdot y}. 
\end{array} \eqno (22) $$

\noindent
where $\xi_{j}$ are the internal variables. Since Eq. (22) is 
rotational invariant, the total orbital angular momentum is
conserved. The wave function $\Psi({\bf x,y})$ with the given total
angular momentum has to belong to an irreducible representation 
of SO($D$). In the traditional method, one calculates the wave 
function by the Clebsch-Gordan coefficients: 
$$\begin{array}{l}
\displaystyle \sum_{l_{D-2},\ldots,l_{1}l'_{D-2},\ldots,l'_{1}}~ 
{\cal Y}^{l}_{l_{D-2},\cdots, l_{1}}({\bf x})
{\cal Y}^{l'}_{l'_{D-2},\cdots, l'_{1}}({\bf y})
\langle l,l_{D-2},\ldots, l_{1};l',l'_{D-2},\ldots,l_{1}|
L,M\rangle. \end{array} \eqno (23) $$

\noindent
As usual, ${\cal Y}^{l}_{l_{D-2},\cdots, l_{1}}({\bf x})$ 
and ${\cal Y}^{l'}_{l'_{D-2},\cdots, l'_{1}}({\bf y})$ are called the
partial angular momentum states, and their combination is
called the total angular momentum state, which is a homogeneous
polynomial of degrees $l$ and $l'$ in the components of ${\bf x}$ 
and ${\bf y}$, respectively. 

There are three problems. First, what kinds of representations 
(or total angular momentum $L$) appear in the Clebsch-Gordan 
series for decomposition of the direct product of two 
representations denoted by one-row Young diagrams $[l,0,\ldots,0]$ 
and $[l',0,\ldots,0]$? This problem has been solved in group theory 
by the Littlewood-Richardson rule and traceless conditions. A new 
character is that the representations denoted by two-row Young 
diagrams appear in the Clebsch-Gordan series for a three-body 
system when $D>3$. Those representations denoted by the Young 
diagrams with more than two rows could not appear because there 
are only two Jacobi coordinate vectors. For simplicity we denote 
a one-row or two-row Young diagram by 
$[\mu,\nu]\equiv [\mu,\nu,0,\ldots,0]$. Hence, we have the 
Clebsch-Gordan series: 
$$[l,0]\otimes [l',0]\simeq 
\displaystyle \bigoplus_{s=0}^{n}\bigoplus_{t=0}^{n-s}~[l+l'-s-2t,s], 
 \eqno (24) $$

\noindent
where $n$ is the minimum between $l$ and $l'$. The representations 
with $t=0$ are calculated by the Littlewood-Richardson rule, 
and the remaining are calculated by the traceless conditions.
The dimension of a representation denoted by a two-row Young
diagram is 
$$\begin{array}{rl}
d_{D}([\mu,\nu])&=~(D+2\mu-2)(D+\mu+\nu-3)(D+2\nu-4)(\mu-\nu+1)\\
&~~~\times\displaystyle {(D+\mu-4)!(D+\nu-5)!\over (\mu+1)!\nu!(D-2)!(D-4)!}.
\end{array} \eqno (25) $$

\noindent
When $D=4$, the representation denoted by a two-row Young 
diagram reduces to a direct sum of a selfdual representation 
$[(S)\mu,\nu]$ and an antiselfdual one $[(A)\mu,\nu]$. Their 
dimensions are equal to half of $d_{4}([\mu,\nu])$ given in 
Eq. (25). When $D=3$, due to the traceless condition, the 
only representations with the two-row Young diagrams are 
representations $[\mu,1]$, which are equivalent to that with 
the one-row Young diagrams $[\mu,0]$, respectively. Eq. (25) 
still holds for $D=3$. Second, how to calculate the Clebsch-Gordan 
coefficients? The calculation must be very complicated. We will avoid the
difficulty by the method of determining the highest weight states
directly. Third, how many base functions are independent for 
a given total orbital angular momentum such that any wave 
function with the same angular momentum can be expanded with 
respect to the base functions where the coefficients are the 
functions of the internal variables. We are going to solve 
the last two problems by group theory. 

Let us sketch some necessary knowledge of group theory. From the 
representation theory of Lie groups \cite{fro,sal,ma2}, the Lie 
algebras of the SO(2$n$+1) group and the SO($2n$) group are $B_{n}$
and $D_{n}$, respectively. Their Chevalley bases with the subscript 
$j$, $1\leq j \leq n-1$, are same: 
$$\begin{array}{l}
H_{j}=L_{(2j-1)(2j)}-L_{(2j+1)(2j+2)},\\
E_{j}=\left(L_{(2j)(2j+1)}-iL_{(2j-1)(2j+1)}
-iL_{(2j)(2j+2)}-L_{(2j-1)(2j+2)}\right)/2, \\
F_{j}=\left(L_{(2j)(2j+1)}+iL_{(2j-1)(2j+1)}
+iL_{(2j)(2j+2)}-L_{(2j-1)(2j+2)}\right)/2. \end{array} \eqno (26a) $$

\noindent
But, the bases with the subscript $n$ are different:
$$\begin{array}{l}
H_{n}=2L_{(2n-1)(2n)},\\
E_{n}=L_{(2n)(2n+1)}-iL_{(2n-1)(2n+1)},\\
F_{n}=L_{(2n)(2n+1)}+iL_{(2n-1)(2n+1)},
\end{array} \eqno (26b) $$

\noindent
for SO($2n+1$), and 
$$\begin{array}{l}
H_{n}=L_{(2n-3)(2n-2)}+L_{(2n-1)(2n)}, \\
E_{n}=\left(L_{(2n-2)(2n-1)}-iL_{(2n-3)(2n-1)}
+iL_{(2n-2)(2n)}+L_{(2n-3)(2n)}\right)/2, \\
F_{n}=\left(L_{(2n-2)(2n-1)}+iL_{(2n-3)(2n-1)}
-iL_{(2n-2)(2n)}+L_{(2n-3)(2n)}\right)/2, \end{array} \eqno (26b) $$

\noindent
for SO($2n$). $H_{k}$ span the Cartan subalgebra, and their eigenvalues 
are the components of a weight vector ${\bf m}=(m_{1},\ldots,m_{n})$: 
$$H_{k}|{\bf m}\rangle = m_{k}|{\bf m}\rangle,~~~~~~1\leq k \leq n. 
\eqno (27) $$

\noindent
If the eigenstates for a given weight ${\bf m}$ are degeneracy,
this weight is called a multiple weight, otherwise a simple one.
$E_{k}$ are called the raising operators and $F_{k}$ the lowering
ones. For an irreducible representation denoted by a Young diagram
$[\mu_{1},\mu_{2},\ldots]$ of SO$(D)$, $\mu_{j}\geq \mu_{j+1}$, 
there is a highest weight ${\bf M}=(M_{1},M_{2},\ldots)$, which 
must be simple:
$$\begin{array}{ll}
M_{j}=\mu_{j}-\mu_{j+1},~~~~~~&1\leq j \leq n-2, \\
M_{n-1}=\mu_{n-1}-\mu_{n},~~~~M_{n}=2\mu_{n},
~~~~&{\rm for~~SO}(2n+1), \\
M_{n-1}=\mu_{n-1}-\mu_{n},~~~~M_{n}=\mu_{n-1}+\mu_{n},
~~&{\rm for~selfdual~representation~in~SO}(2n), \\
M_{n-1}=\mu_{n-1}+\mu_{n},~~~~M_{n}=\mu_{n-1}-\mu_{n},
~~&{\rm for~antiselfdual~representation~in~SO}(2n).
\end{array} \eqno (28) $$

\noindent
We are not interested here in the spinor representations where 
$M_{n}$ is odd for SO($2n+1$) and $M_{n-1}+M_{n}$ is odd for
SO($2n$). For a given irreducible representation 
$[\mu_{1},\mu_{2},\ldots]$ of SO$(D)$, we only need to consider 
the highest weight state $|{\bf M}\rangle$, which satisfies
$$H_{k}|{\bf M}\rangle=M_{k}|{\bf M}\rangle,~~~~~~
E_{k}|{\bf M}\rangle=0,~~~~~~1\leq k \leq n, \eqno (29) $$

\noindent
because its partners can be calculated by the lowering operators $F_{k}$. 
In this paper the highest weight state will simply be called the 
wave functions with the given angular momentum $[\mu,\nu]$ for 
simplicity. 

Now, we return to our problems. Recalling the Clebsch-Gordan series
in Eq. (24), we can rewrite Eq. (23) for the highest weight state 
${\bf M}$:
$$\begin{array}{l}
{\cal Y}^{l,l',s,t}_{\bf M}({\bf x,y})=
\displaystyle \sum_{\bf m}~
{\cal Y}^{l}_{\bf m}({\bf x})
{\cal Y}^{l'}_{\bf M-m}({\bf y})
\langle l,{\bf m}, l', ({\bf M-m})~|~[(l+l'-s-2t),s],{\bf M}\rangle, 
\end{array} \eqno (30) $$

\noindent
where the subscripts of the harmonic polynomials are changed to
the weights for simplicity. 
${\cal Y}^{l,l',s,t}_{\bf M}({\bf x,y})$ is the highest weight 
state of the representation $[(l+l'-s-2t),s]$. It is a homogeneous 
polynomial of degrees $l$ and $l'$ in the components 
of ${\bf x}$ and ${\bf y}$, respectively. Generally speaking, 
some ${\cal Y}^{l,l',s,t}_{\bf M}({\bf x,y})$ may be expressed as a
sum where each term is a product of an internal variable $\xi_{j}$ 
and a homogeneous polynomial $f({\bf x,y})$ of lower degree (see 
p. 042108-5 in \cite{gu}). Since ${\cal Y}^{l,l',s,t}_{\bf M}({\bf x,y})$ 
will be used as a base function for the wave function with a given 
angular momentum and the combinative coefficient is the function 
of the internal variables, in this meaning, the base function 
in the form of $\xi_{j}f({\bf x,y})$ is not independent, and 
we should find out the independent and complete base functions 
for any given angular momentum $[\mu,\nu]$. In the following 
we are going to prove 
${\cal Y}^{q,(\mu+\nu-q),\nu,0}_{\bf M}({\bf x,y})$ 
and their partners, where $l=q$, $l'=\mu+\nu-q$, $s=\nu$, $t=0$, 
and $\nu\leq q \leq \mu$, constitute a complete set of 
independent base functions for the total orbital angular momentum 
$[\mu,\nu]$. In other words, those total angular momentum states
${\cal Y}^{l,l',s,t}_{\bf M}({\bf x,y})$ with $t>0$ are not independent,
where the sum of the partial angular momentum quantum number 
$l$ and $l'$ is larger than $\mu+\nu$ for the total angular momentum
$[\mu,\nu]$. 

The highest weight for the representation $[\mu,\nu]$ is 
${\bf M}=(\mu-\nu,\nu,0,\ldots,0)$. Removing the normalization 
factor in ${\cal Y}^{q,(\mu+\nu-q),\nu,0}_{\bf M}({\bf x,y})$, 
which is irrelevant here, we can determine the explicit form for 
${\cal Y}^{q,(\mu+\nu-q),\nu,0}_{\bf M}({\bf x,y})$ according to
its orders in the components of ${\bf x}$ and ${\bf y}$ and 
the property of the highest weight state (29), and denote it
by the generalized harmonic polynomial $Q^{\mu \nu}_{q}({\bf x,y})$:
$$\begin{array}{rl}
Q^{\mu \nu}_{q}({\bf x,y})&=~\displaystyle {X_{12}^{q-\nu}Y_{12}^{\mu-q}
\over (q-\nu)!(\mu-q)!}
\left(X_{12}Y_{34}-Y_{12}X_{34}\right)^{\nu}\\[2mm]
&\sim~ {\cal Y}^{q,(\mu+\nu-q),\nu,0}_{\bf M}({\bf x,y}),~~~~~~
0\leq \nu \leq q \leq \mu,\\
X_{12}=x_{1}+ix_{2},~~~&X_{34}=x_{3}+ix_{4},~~~~~~
Y_{12}=y_{1}+iy_{2},~~~~~~Y_{34}=y_{3}+iy_{4}.
\end{array} \eqno (31) $$

\noindent
The formula for $Q^{\mu \nu}_{q}({\bf x,y})$ holds for $D=3$ 
($x_{4}=y_{4}=0$, $\nu=0$ or $1$) \cite{ma1,gu} and $D>4$. When 
$D=4$ we denote the highest weight states by 
$Q^{(S)\mu \nu}_{q}({\bf x,y})$ and $Q^{(A)\mu \nu}_{q}({\bf x,y})$ 
for the selfdual representations and the antiselfdual representations, 
respectively:
$$\begin{array}{c}
Q^{(S)\mu \nu}_{q}({\bf x,y})=\displaystyle {X_{12}^{q-\nu}Y_{12}^{\mu-q}
\over (q-\nu)!(\mu-q)!}
\left(X_{12}Y_{34}-Y_{12}X_{34}\right)^{\nu}\\[2mm]
Q^{(A)\mu \nu}_{q}({\bf x,y})=\displaystyle {X_{12}^{q-\nu}Y_{12}^{\mu-q}
\over (q-\nu)!(\mu-q)!}
\left(X_{12}Y_{34}^{\prime}-Y_{12}X_{34}^{\prime}\right)^{\nu}\\[2mm]
X^{\prime}_{34}=x_{3}-ix_{4},~~~~~~Y^{\prime}_{34}=y_{3}-iy_{4}.
\end{array} \eqno (32) $$

\noindent
The generalized harmonic polynomial $Q^{\mu \nu}_{q}({\bf x,y})$
is a homogeneous polynomial of degrees $q$ and $(\mu+\nu-q)$
in the components of ${\bf x}$ and ${\bf y}$, respectively. It is 
a simultaneous eigenfunction of 
$\bigtriangledown^{2}_{\bf x}$, $\bigtriangledown^{2}_{\bf y}$, 
$\bigtriangledown_{\bf x} \cdot \bigtriangledown_{\bf y}$, and 
the total angular momentum operator ${\bf L}^{2}$ [see Eq. (9)],
$$\begin{array}{l}
\bigtriangledown^{2}_{\bf x}Q^{\mu \nu}_{q}({\bf x,y})
=\bigtriangledown^{2}_{\bf y}Q^{\mu \nu}_{q}({\bf x,y})
=\bigtriangledown_{\bf x}\cdot
\bigtriangledown_{\bf y}Q^{\mu \nu}_{q}({\bf x,y})=0,\\
{\bf L}^{2}Q^{\mu \nu}_{q}({\bf x,y})=C_{2}([\mu,\nu])
Q^{\mu \nu}_{q}({\bf x,y}),\\
C_{2}([\mu,\nu])=\mu(\mu+D-2)+\nu(\nu+D-4),
\end{array} \eqno (33) $$

\noindent
where $C_{2}([\mu ,\nu])$ is the Casimir calculated by a general 
formula (see (1.131) in Ref. \cite{ma2}). The parity of 
$Q^{\mu \nu}_{q}({\bf x,y})$ is obviously $(-1)^{\mu+\nu}$. 

It is evident that $Q^{\mu \nu}_{q}({\bf x,y})$ do not contain
a function of the internal variables as a factor, neither do
their partners due to the rotational symmetry. Therefore, 
$Q^{\mu \nu}_{q}({\bf x,y})$ are independent base functions 
for the given angular momentum described by $[\mu,\nu]$. 
Now, we are going to prove that $(\mu-\nu+1)$ base functions
$Q^{\mu \nu}_{q}({\bf x,y})$ where $\nu\leq q \leq \mu$ are complete
for the angular momentum $[\mu,\nu]$. That is, 
$Q^{\mu(l-\mu)}_{q}({\bf x,y})$ with $0\leq l-\mu \leq q \leq \mu$ 
and their partners construct a complete set of linearly independent
base functions for the homogeneous polynomials of degree $l$ 
in the components of ${\bf x}$ and ${\bf y}$, excluded those in
the forms of $\xi_{j}f({\bf x,y})$, where $f({\bf x,y})$ is a 
homogeneous polynomial of degree $(l-2)$.

The number of linearly independent homogeneous polynomials of 
degree $l$ in the components of ${\bf x}$ and ${\bf y}$ is 
$$M_{D}(l)=\left(\begin{array}{c} l+2D-1 \\ 2D-1 \end{array} \right).  $$

\noindent
After removing those polynomials in the form $\xi_{j}f({\bf x,y})$, 
the number $M(l)$ reduces to $K(l)$:
$$\begin{array}{rl}
K_{D}(l)&=~M_{D}(l)-3M_{D}(l-2)+3M_{D}(l-4)-M_{D}(l-6)\\
&=4(l+D-3)\left[2l(l+2D-6)+(D-2)(2D-5)\right]
\displaystyle {(l+2D-7)! \over l!(2D-4)!},
\end{array} \eqno (34) $$

\noindent
when $l+2D\geq 7$, which only excludes one case of $l=0$ and $D=3$, 
where $K_{3}(0)=1$. 

On the other hand, the number of $Q^{\mu(l-\mu)}_{q}({\bf x,y})$ with 
$0\leq l-\mu \leq q \leq \mu$ and their partners can be calculated 
directly from Eq. (25):
$$\displaystyle \sum_{l/2\leq \mu \leq l}~(2\mu-l+1)d_{D}([\mu,(l-\mu)])
=K_{D}(l). \eqno (35) $$

\noindent
Eqs. (34) and (35) are checked by Mathematica. Thus, we have proved 
that $(\mu-\nu+1)$ base functions $Q^{\mu \nu}_{q}({\bf x,y})$ 
where $0\leq \nu\leq q \leq \mu$ are independent and complete 
for the angular momentum $[\mu,\nu]$. Any function with 
the angular momentum $[\mu,\nu]$ in the system can be expanded 
with respect to the base functions $Q^{\mu \nu}_{q}({\bf x,y})$,
where the coefficients are functions of internal variables.  

From Eq. (30), for a given total orbital angular momentum 
$[\mu,\nu]$ there are infinite number of wave functions 
${\cal Y}^{(q+t),(\mu+\nu+t-q),\nu,t}_{\bf M}({\bf x,y})$ 
combined from different partial angular momentum states. Now, 
what we have proved is that only a finite number of partial angular 
momentum states ($t=0$) are involved in the complete set of 
independent base functions $Q^{\mu\nu}_{q}({\bf x,y})$ for a 
given total orbital angular momentum $[\mu,\nu]$. 

\section{Generalized radial equations}

In the preceding section we proved that any function with angular 
momentum $[\mu,\nu]$ in the quantum three-body system of $D$ dimensions 
can be expanded with respect to the base functions 
$Q^{\mu \nu}_{q}({\bf x,y})$
$$\Psi^{[\mu,\nu]}_{\bf M}({\bf x,y})=\displaystyle \sum_{q=\nu}^{\mu}~
\psi^{\mu \nu}_{q}(\xi_{1},\xi_{2},\xi_{3})
Q^{\mu \nu}_{q}({\bf x,y}), \eqno (36) $$

\noindent
where the coefficients $\psi^{\mu \nu}_{q}(\xi_{1},\xi_{2},\xi_{3})$
are called the generalized radial functions. When substituting Eq. (36)
into the Schr\"{o}dinger equation (22), the main calculation in the 
derivation is to apply the Laplace operator (21) to the function 
$\Psi^{[\mu,\nu]}_{\bf M}({\bf x,y})$. The calculation consists of three 
parts. The first is to apply the Laplace operator to the generalized 
radial functions $\psi^{\mu \nu}_{q}(\xi_{1},\xi_{2},\xi_{3})$,
which can be calculated by replacement of variables: 
$$\begin{array}{l}
\bigtriangledown^{2} \psi^{\mu \nu}_{q}(\xi_{1},\xi_{2},\xi_{3})=
\left\{ 4\xi_{1}\partial^{2}_{\xi_{1}}+4\xi_{2}\partial^{2}_{\xi_{2}}
+2 D\left(\partial_{\xi_{1}}+\partial_{\xi_{2}}\right)
+\left(\xi_{1}+\xi_{2}\right)\partial^{2}_{\xi_{3}}\right.\\
\left.~~~~~~~~~~~+4\xi_{3}\left(\partial_{\xi_{1}}+\partial_{\xi_{2}}\right)
\partial_{\xi_{3}}\right\}\psi^{\mu \nu}_{q}(\xi_{1},\xi_{2},\xi_{3}),
\end{array} \eqno (37) $$

\noindent
where $\partial_{\xi}$ denotes $\partial/\partial\xi$ and so on. 
The second is to apply it to the generalized harmonic polynomials 
$Q_{q}^{\mu \nu}({\bf x,y})$. This part is vanishing because 
$Q_{q}^{\mu \nu}({\bf x,y})$ satisfies the Laplace equation. 
The last is the mixed application
$$\begin{array}{l}
2\left\{\left(\partial_{\xi_{1}}\psi^{\mu \nu}_{q}\right)2{\bf x}
+\left(\partial_{\xi_{3}}\psi^{\mu \nu}_{q}\right){\bf y}\right\} 
\cdot \bigtriangledown_{\bf x} Q_{q}^{\mu \nu}
+2\left\{\left(\partial_{\xi_{2}}\psi^{\mu \nu}_{q}\right)2{\bf y}
+\left(\partial_{\xi_{3}}\psi^{\mu \nu}_{q}\right){\bf x}\right\} 
\cdot \bigtriangledown_{\bf y} Q_{q}^{\mu \nu}. 
\end{array} \eqno (38) $$

\noindent
From the definition (31) for $Q_{q}^{\mu \nu}({\bf x,y})$ we
have
$$\begin{array}{ll}
{\bf x}\cdot \bigtriangledown_{\bf x}Q_{q}^{\mu \nu}
=qQ_{q}^{\mu \nu},~~~~~~
&{\bf y}\cdot \bigtriangledown_{\bf y}Q_{q}^{\mu \nu}
=(\mu+\nu-q)Q_{q}^{\mu \nu}\\
{\bf y}\cdot \bigtriangledown_{\bf x}Q_{q}^{\mu \nu}
=(\mu-q+1)Q_{q-1}^{\mu \nu},~~~~~~
&{\bf x}\cdot \bigtriangledown_{\bf y}Q_{q}^{\mu \nu}
=(q-\nu+1)Q_{q+1}^{\mu \nu}. \end{array} \eqno (39) $$

\noindent
Hence, we obtain the generalized radial equation, satisfied
by the $(\mu-\nu+1)$ functions $\psi^{\mu\nu}_{q}(\xi_{1},\xi_{2},\xi_{3})$:
$$\begin{array}{c}
\bigtriangledown^{2} \psi^{\mu \nu}_{q} +4q \partial_{\xi_{1}}
\psi^{\mu \nu}_{q} +4(\mu+\nu-q) \partial_{\xi_{2}}
\psi^{\mu \nu}_{q}
+2(\mu-q) \partial_{\xi_{3}}\psi^{\mu \nu}_{q+1} 
+2(q-\nu)  \partial_{\xi_{3}}\psi^{\mu \nu}_{q-1}\\
=-2\left(E-V\right) \psi^{\mu \nu}_{q},
\end{array} \eqno (40) $$

\noindent
where $\bigtriangledown^{2} \psi^{\mu \nu}_{q}$ is given in Eq. (37). 
Only three invariant variables $\xi_{1}$, $\xi_{2}$ and $\xi_{3}$
are involved both in the equations and in the functions. When $D=4$, 
Eq. (40) holds for the generalized radial functions either in 
$[(S)\mu,\nu]$ or in $[(A)\mu,\nu]$, because two representations
incorporate to one irreducible representation of the O(4) group 
when the space inversion is considered. When $D=3$, the equations 
for the functions in $[\mu,0]$ and in $[\mu,1]$ are different 
although two representations $[\mu,0]$ and $[\mu,1]$ are equivalent, 
because the functions have different parity.   

At last, we discuss rotational variables and the volume
element of the configuration space. We fix the body-fixed frame
such that ${\bf x}$ is parallel with its $D$th axis, and ${\bf y}$
is located in its $(D-1)D$ hyperplane with a non-negative 
$(D-1)$th component. That is, in the body-fixed frame, the nonvanishing
components of two Jacobi coordinate vectors ${\bf x}'$ and
${\bf y}'$ are
$$x'_{D}=\xi_{1}^{1/2},~~~~
y'_{D-1}=\left(\xi_{2}-\xi_{3}^{2}/\xi_{1}\right)^{1/2},~~~~
y'_{D}=\xi_{3}\xi_{1}^{-1/2}. \eqno (41) $$

\noindent
Let $R=R^{(1)}R^{(2)}\in$SO($D$) rotate the center-of-mass frame to
the body-fixed frame: 
$$\begin{array}{l}
R^{(1)}=R_{12}(\theta_{1})R_{31}(\theta_{2})R_{43}(\theta_{3})
R_{54}(\theta_{4})\ldots R_{D(D-1)}(\theta_{D-1}), \\
R^{(2)}=R_{12}(\varphi_{1})R_{31}(\varphi_{2})R_{43}(\varphi_{3})
R_{54}(\varphi_{4})\ldots R_{(D-1)(D-2)}(\varphi_{D-2}), \\
R{\bf x}'={\bf x},~~~~~~R{\bf y}'={\bf y},  \end{array} \eqno (42) $$

\noindent
where, for example, $R_{12}(\theta)$ is a rotation on the hyperplane
with the first and the second axes through $\theta$ angle: 
$$R_{12}(\theta)=\left(\begin{array}{ccc} \cos \theta & -\sin \theta & 0\\ 
\sin \theta & \cos \theta & 0 \\ 0 & 0 & {\bf 1}_{D-2}\end{array} \right). $$

\noindent
$(D-1)$ $\theta_{j}$ and $(D-2)$ $\varphi_{k}$ are the rotational
variables, called the generalized Euler angles. Through a straightforward
calculation, we obtain 
$$\begin{array}{l}
x_{1}+ix_{2}=\xi_{1}^{1/2}e^{i\theta_{1}}
\displaystyle \prod_{a=2}^{D-1}~\sin \theta_{a},~~~~~~
x_{3}+ix_{4}=\xi_{1}^{1/2}\left(\cos\theta_{2}\sin\theta_{3}
+i\cos\theta_{3}\right)
\displaystyle \prod_{a=4}^{D-1}~\sin \theta_{a},\\
y_{1}+iy_{2}=\xi_{3}\xi_{1}^{-1/2}e^{i\theta_{1}}
\displaystyle \prod_{a=2}^{D-1}~\sin \theta_{a}
+\left(\xi_{2}-\xi_{3}^{2}/\xi_{1}\right)^{1/2}e^{i\theta_{1}}\\
~~~\times\left\{
i\displaystyle \prod_{a=1}^{D-2}~\sin \varphi_{a}
+\displaystyle \sum_{a=1}^{D-2}~\cos\theta_{a+1}\cos\varphi_{a}
\left(\displaystyle \prod_{b=2}^{a}~\sin\theta_{b}\right)
\left(\displaystyle \prod_{c=a+1}^{D-2}~\sin\varphi_{c}\right)\right\},\\
\end{array} $$ 
$$\begin{array}{l}
y_{3}+iy_{4}=\xi_{3}\xi_{1}^{-1/2}\left(\cos\theta_{2}\sin\theta_{3}
+i\cos\theta_{3}\right)\displaystyle \prod_{a=4}^{D-1}~\sin \theta_{a}
+\left(\xi_{2}-\xi_{3}^{2}/\xi_{1}\right)^{1/2}\\
~~~\times~\left\{
-\cos\varphi_{1}\sin\theta_{2}
\displaystyle \prod_{a=2}^{D-2}~\sin \varphi_{a}
+\left(\cos \theta_{2}\cos\theta_{3}-i\sin\theta_{3}\right)\cos\varphi_{2}
\displaystyle \prod_{a=3}^{D-2}~\sin \varphi_{a}\right.\\
\left.~~~+\left(\cos\theta_{2}\sin\theta_{3}+i\cos\theta_{3}\right)
\displaystyle \sum_{a=3}^{D-2}~\cos\theta_{a+1}\cos\varphi_{a}
\left(\displaystyle \prod_{b=4}^{a}~\sin\theta_{b}\right)
\left(\displaystyle \prod_{c=a+1}^{D-2}~\sin\varphi_{c}\right)\right\},
\end{array} \eqno (43) $$

\noindent
where $\prod_{a=b+1}^{b}F_{a}=1$. The volume element of the 
configuration space is 
$$\displaystyle \prod_{j=1}^{D}~dx_{j}dy_{j}
=\displaystyle {1\over 4}\left(\xi_{1}\xi_{2}-\xi_{3}^{2}\right)^{(D-3)/2}
d\xi_{1}d\xi_{2}d\xi_{3}
\displaystyle \prod_{j=1}^{D-1}~
\left(\sin\theta_{j}\right)^{j-1}d\theta_{j}
\displaystyle \prod_{k=1}^{D-2}~
\left(\sin\varphi_{k}\right)^{k-1}d\varphi_{k}. \eqno (44) $$

\section{Conclusions}

After separating the motion of center of mass, we have defined
the homogeneous polynomial $Q^{\mu\nu}_{q}({\bf x,y})$ of degree 
$q$ and $(\mu+\nu-q)$ in the components of the Jacobi coordinate 
vectors ${\bf x}$ and ${\bf y}$, respectively. $Q^{\mu\nu}_{q}({\bf x,y})$
is a solution of the Laplace equation. We have proved that 
$(\mu-\nu+1)$ generalized harmonic polynomials 
$Q^{\mu\nu}_{q}({\bf x,y})$ constitute a complete 
set of independent base functions for the total orbital angular 
momentum $[\mu,\nu]$. Any wave function with the given angular 
momentum in the system can be expanded with respect to the base 
functions, where the coefficients are the functions of the 
internal variables, called the generalized radial functions. The 
three-body Schr\"{o}dinger equation with a spherically symmetric 
potential $V$ in $D$ dimensions reduces to the generalized radial 
equations satisfied by the generalized radial functions. Only three 
internal variables are involved in the functions and equations. 
The number of both the functions and the equations for the 
given angular momentum $[\mu,\nu]$ is finite and equal to 
$\left(\mu-\nu+1\right)$. Only a finite number of partial angular 
momentum states are involved in constructing the generalized 
harmonic polynomials $Q^{\mu\nu}_{q}({\bf x,y})$, and the 
contributions from the remaining partial angular momentum states 
have been incorporated into those from the generalized radial functions. 

The generalization of this method to a quantum $N$-body system
in $D$-dimensions is straightforward. The difficulty is how to 
obtain the unified forms for the generalized harmonic polynomials, 
because it needs $D-1$ vectors to determine the body-fixed frame 
and there are $N-1$ Jacobi coordinate vectors. The cases with 
$N<D$ are very different to the cases with $N\geq D$. We will 
study this problem elsewhere. 

\noindent
{\bf ACKNOWLEDGMENTS}This work is supported by the 
National Natural Science Foundation of China.

\end{document}